\documentclass{elsart3}
\usepackage{epsfig}
\usepackage{amssymb}
\begin{document}
\begin{frontmatter}

\title{SHAPE OF A BARKHAUSEN PULSE}

\author[1]{F. Colaiori}
\author[1]{S. Zapperi}
\author[2]{G. Durin}
\address[1]{INFM-SMC, Dipartimento di Fisica,
Universit\`a "La Sapienza", P.le A. Moro 2
00185 Roma, Italy}
\address[2]{Istituto Elettrotecnico Nazionale Galileo Ferraris, 
strada delle Cacce 91, I-10135 Torino, Italy}
\begin{abstract}
The average shape of the pulse in Barkhausen noise has been recently proposed 
as a tool to compare models and experiments. We compute theoretically the 
pulse shape of Barkhausen noise in a model describing 
the motion of a domain wall in an effective Brownian potential. In this framework, 
the pulse shape is related to the properties of the excursion of a random process 
in a $c\log(x)-kx$ potential. We record the Barkhausen noise in polycristalline 
$\mbox{FeSi}$ materials, and  compare the pulse shape  
with the one predicted by the domain wall model. 
\end{abstract}

\end{frontmatter}

The Barkhausen noise has been incessantly investigated because of its 
practical application and theoretical implications. 
Experiments show that both size and duration of avalanches of spin 
reversal are power law distributed over several decades. 
Recently, the average pulse shape has been proposed as a sharper tool 
to test models against experiments \cite{KUN-00}. 

The scaling analysis suggested in Ref.~\cite{KUN-00} is based on a simple relation
between the average size $\langle S \rangle$ of an avalanche and its duration $T$
scaling as $\langle S \rangle \sim T^{1/\sigma \nu z}$ 
where $1/\sigma \nu z$ is a combination of critical exponents, defined in
Ref.~\cite{KUN-00}. The average avalanche has an universal shape
given by $v(t, T) \sim T^{1/\sigma \nu z-1} g(t/T)$,
where $g$ is a universal scaling function. Similarly, when considering 
the signal $v$ as a function of the magnetization $s = \int_0^t v dt$, 
one gets $v(s, S) \sim S^{1-1/\sigma \nu z} f(s/S)$
where  $f$ is another scaling function. 
In the model presented in Ref.~\cite{KUN-00}, $g$ is found to be
an inverted parabola, but this result does not fit with experiments.

The Barkhausen noise can be described in terms of a phenomenological model, 
known as ABBM \cite{ALE-90}, which describes the wall 
as a rigid interface in an effective Brownian pinning potential. The crucial assumption 
of a Brownian correlated pinning potential was done on phenomenological basis, since this 
kind of correlation is observed in experiments, but the model can be obtained
as a mean-field version of a more general flexible domain wall model \cite{ZAP-98}.
Considering the  domain wall velocity as a function of magnetization $s$, one obtains the 
following Langevin equation \cite{ZAP-98}
\begin{equation}
\frac{dv}{ds}= \frac{c}{v}-k+\eta(s) \label{eq:rw}
\end{equation}
where $\eta$ is uncorrelated Gaussian noise with $\langle \eta(s) \eta(s')\rangle=2
\delta(s-s')$, and $c$ is a dimensionless parameter proportional to the applied field
rate. If we neglect the contribution of the demagnetizing factor $k$, Eq.~\ref{eq:rw}
describes the motion of a 1$d$ random walk in a logarithmic potential $E(v)=-c
\,\log(v)$, where the magnetization $s$ plays the role of time. The corresponding
Fokker--Plank equation is
\begin{equation}
\frac{\partial P(v,s)}{\partial s}= \frac{\partial}{\partial v} \left(
-\frac{c}{v}+\frac{\partial}{\partial v} P(v,s) \right),
\end{equation}
where $P(v,s)$ is the probability to find the walk in $v$ at $s$. We are interested in
a solution of this equation with the initial condition $P(v,0)=\delta(v-v_0)$ and an
absorbing boundary at the origin $P(v=0,t)=0$. 
This solution can be expressed in terms of modified Bessel functions \cite{BRA-00}. 
For the interesting case $0<c<1$, which is the condition of the ABBM model to have power 
laws in the avalanche distribution \cite{ALE-90}, the probability $P(v,s | v_0, 0;c)$ for 
a walk starting at
$v_0$ to be at $v$ after a ``time'' $s$, in the limit $v_0 \rightarrow 0$ is simply
proportional to a power of $v$ times a Gaussian with variance $s$, and the average
excursion is the ratio between two successive moments of a Gaussian with variance
$w=2s(S-s)$, and it is thus simply proportional to $\sqrt{w}$. The normalized
average excursion is therefore given by
$\langle v \rangle \propto \sqrt{s(S-s)}$. 
It is also possible to calculate the function $g$. By
definition the avalanche size $s$ at time $t$ is given by the integral of $v(t,T)$ from
time zero to time $t$:
\begin{equation}
s=\int_0^t dt' v(t',T)\propto T^{1/\sigma \nu z}\int_0^{t/T} g(x)dx \,,
\end{equation}
which provides an expression of $s=s(t,T)$ as a function of $t$ and $T$.
Imposing
$v(t,T)=v(s=s(t,T),S(T))$
gives an integral equation for $g$  involving $f$: 
$g(x) \propto f\left(\int_0^{x} g(x')dx'\right)$. Using the form of $f$, we can solve this
equation with the boundary conditions $g(0)=g(1)=1$. The solution is
$g(x) \propto \sin(\pi x)$.  
Summarizing, for the normalized avalanche we obtain
\begin{equation}
v(s,S)=S^{1-\sigma \nu z} \pi\sqrt{(s/S)(1-s/S)} \,,\label{eq:Vtnorm}
\end{equation}
\begin{equation}
\label{eq:Vsnorm} v(t,T)=T^{1/\sigma \nu z -1} \pi/2\sin(\pi t/T) \,.
\end{equation}

\begin{figure}
\includegraphics[height=8cm]{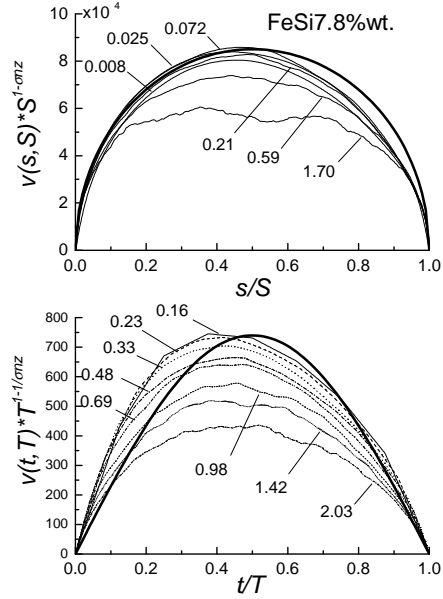} 
\caption   { \label{fig:semi} Average avalanche shapes for
polycrystalline FeSi material. Time signal $v(t,T)$
and magnetization signal $v(s,S)$ are scaled according to the text.
Bold lines are the theoretical predictions. 
Numbers in the graphs denote avalanche size in Wb (top), 
and duration in sec (bottom).}
\end{figure}

It is interesting to compare the theoretical results given above with the experimental
average shapes. In
Figs.~\ref{fig:semi} we plot the signal voltages both as a function of time $v(t,T)$
and magnetization $v(s,S)$ rescaled using the theoretical value, $1/\sigma \nu z =2$.
The demagnetizing fields correspond to a bias to the random walk, which has the effect 
of introducing a cutoff in the distribution of avalanche sizes and durations, but 
has no effect on the shape of the scaling functions. Despite the asymmetry and the 
deviation from scaling observed in the experimental shapes, the predictions based 
on the ABBM model are in reasonable agreement with experiments, 
supporting the validity of the interface model approach.

\end{document}